# Replacing Gallium Oxide shell with conductive Ag: Toward printable and recyclable composite for highly stretchable electronics, electromagnetic shielding, and thermal interfaces

Abdollah Hajalilou[1], Elahe Parvini[1], Tiago A. Morgado[2], Pedro Alhais Lopes[1], M. Estrela Melo Jorge[3] , Marta Freitas[1], Mahmoud Tavakoli[1,*]

1. Institute of Systems and Robotics, Department of Electrical Engineering, University of Coimbra, Coimbra 3030-290, Portugal.

2. Instituto de Telecomunicações and Department of Electrical Engineering, University of Coimbra, 3030-290 Coimbra, Portugal.

3. Biosystems and Integrative Sciences Institute (BioISI), Faculdade de Ciências, Universidade de Lisboa, 1749-016 Lisboa, Portugal.

[*] Crossponding E-mail: mahmoud@isr.uc.pt

**Abstract**

Liquid metal (LM)-based composites hold promise for soft electronics due to their high conductivity and fluidic nature. However, the presence of $\alpha$-$Ga_2O_3$ and GaOOH layers around LM droplets impairs conductivity and performance. We tackle this issue by replacing the oxide layer with conductive silver (Ag) using an ultrasonic-assisted galvanic replacement reaction. The Ag-coated nanoparticles form aggregated, porous microparticles that are mixed with Styrene-Isoprene-Styrene (SIS) polymers, resulting in a digitally printable composite with superior electrical conductivity and electromechanical properties compared to conventional fillers. Adding more LM enhances these properties further. The composite achieves EMI shielding effectiveness (SE) exceeding 75 dB in the X-band frequency range, even at 200% strain, meeting stringent military and medical standards. It is applicable in wireless communications, Bluetooth signal blocking, and as a thermal interface material (TIM). Additionally, we highlight its recyclability using a biodegradable solvent, underscoring its eco-friendly potential. This composite represents a significant advancement in stretchable electronics and EMI shielding, with implications for wearable and bioelectronic applications.

**Keywords:** Stretchable Liquid Metal Composites; Intermetallic Filler Particles; Digital printing, Electromagnetic Interference (EMI) Shielding; and Thermal Interface Materials; Recyclable Electronics; Galvanic Replacement; Soft Electronic.

## Introduction

The increasing reliance on electronic devices has led to a surge in electromagnetic interference (EMI), posing challenges to device functionality and human health [1-3]. Traditional EMI shielding materials are often rigid, limiting their applicability in emerging soft electronics and wearable technologies [4-5]. A shift toward soft and stretchable EMI shielding materials is crucial for emerging applications like body-interfacing conformal electronics, wearable patches, and bioelectronics, demanding innovative materials with stretchability, conductivity, self-healing properties, and recyclability [6-8]. Metal films and metal-based composites have been utilized for effective reflection-based shielding due to their high conductivity [9-10]. Among them are particle-filled conductive composites with metallic or carbon-based fillers [11-12]. However, challenges remain in achieving high SE under mechanical strain. Increasing the particle filler loads in polymer composites generally improves both conductivity and EMI shielding efficacy, but limits processability and flexibility, and results in increased stiffness and brittleness [13-14].

Gallium-based liquid metal (LM) alloys have recently emerged as promising candidates in soft electronic applications due to their intrinsic conductivity and fluidic nature [15-17]. However, LMs are challenging to process and deposit/print, exhibit limited adherence to the substrate [16], and are unstable against mechanical touch due to their liquid phase and smearing nature. Moreover, these LM materials alone do not exhibit the desired electromagnetic shielding behavior, which we will discuss in the electromagnetic interference (EMI) shielding section. Furthermore, when liquid Ga is mixed with thermoplastic polymers such as Styrene-Isoprene-Styrene (SIS), the resulting composite with nonconductive behaviour conductivity and requires further treatment, such as mechanical sintering, to become conductive [18-19]. Another critical challenge associated with these LMs is the formation of native oxide films, primarily composed of gallium oxide ($Ga_2O_3$) and gallium oxyhydroxide (GaOOH), when exposed to air [20-21]. These oxide films are semiconducting and disrupt the electrical connectivity between liquid metal droplets, leading to reduced electrical and electromechanical performance, which limits their application in stretchable circuits and wearable, and soft electronics [22-23].

On the other hand, the oxide layers formed on the Ga LM droplets present an opportunity to create other compounds over the core Ga. In this context, the galvanic replacement reaction (GRR) has been identified as an effective solution [24-25]. This method involves replacing the oxide layer with a noble metal, such as silver (Ag), gold (Au), or platinum (Pt), which offers superior electrical conductivity and resistance to further oxidation. Silver is particularly advantageous due to its high electrical conductivity, helping to restore and enhance the electrical pathways disrupted by the oxide layer. This process takes advantage of the significant difference in standard redox potentials between gallium (Ga) and noble metals like silver, gold, or platinum. For example, the standard reduction potential of Ga ($Ga^{3+}/Ga$) is much lower than that of Ag ($Ag^+/Ag$), making the spontaneous replacement of the oxide layer by silver both feasible and effective. The silver coating restores and enhances electrical connectivity, improves biocompatibility, and facilitates the broader application of liquid metals in advanced technologies. This approach not only mitigates the limitations imposed by the oxide layer but also leverages the benefits of noble metals to enhance the overall performance and versatility of liquid metal-based systems.

In this work, we developed a biphasic composite synthesized by surface modification of Ga LM particles with an aggregated Ag shell through ultrasonic galvanic replacement. This approach enables the synthesis of diverse complex nanostructured materials with adjustable shapes, sizes, and distributions, including intermetallic compounds and particles with sophisticated surface morphologies [26-27]. Tuning parameters result in multimetallic nanomaterials with enhanced adjustable performance, as the electrochemical replacement of Ga LM atoms by metals influences the final product's composition and morphology [25, 28-29]. Such strategies— the ultrasonic galvanic replacement technique— address challenges related to viscosity and processing. It can also be used to tune the electrical properties of the composite. Consequently, these strategies suggest the LMs are more versatile for achieving effective EMI shielding due to the effect of interfacial morphology on EMI shielding performance.

In this study, we aim to study the formation of new biphasic composites based on ultrasonic galvanic replacement reactions (GRRs) and target the optimization of synthesis parameters to obtain a composite that can be applied over various surfaces through simple deposition techniques such as stencil printing, and digital printing. We particularly show Ag-coated liquid metal nanodroplets that are effective in EMI shielding, and can maintain the shielding under mechanical strain. Such shielding films are especially attractive for wearable devices to protect human health against the EM field. We also show that aggregates of Ag-coated LM droplets can form microparticles for conductive composites that present advantages over previous particle-filled conductive composites. Finally, we demonstrate the recyclability of the developed composite incorporating intermetallic-coated liquid metal particles, through processing with a biodegradable deep eutectic solvent (DES).

**Figure 1** illustrates the composite fabrication process, from the use of as-synthesized Ag-Ga intermetallic filler particles to characterization, processing, applications, and recycling. In this study, we perform ultrasonic galvanic replacement reactions (GRRs) to synthesize the Ga@Ag core-shell structure in the presence of molten Ga and a silver nitrate solution. At low concentrations of silver nitrate (0.1 mM to 5 mM), along with Ag nanoparticles, we observed the formation of undesired Ga oxides, i.e., $\alpha$-$Ga_2O_3$ and GaOOH, on the surface of Ga particles. When the molar ratio of silver nitrate increases to 0.5 M, the desirable product forms—Ga coated with biphasic $GaAg_2$ alongside Ag nanoparticles on the surface of Ga particles—forms. Furthermore, with the increase in the molar ratio from millimolar to molar in silver nitrate concentration, we transit from pasty-like particles to a microstructure is composed of aggregated biphasic droplets; bringing advantages both for conductive composites and EM shielding films. In one experiment, we fabricated a conductive composite by blending aggregated microparticles with a Styrene-Isoprene-Styrene (SIS) polymer composite as well as with EGaIn liquid metal (**Figure 1A**) and analyzed its electrical and electromechanical properties. The as-fabricated composites were subsequently characterized through S-parameter measurements performed with a vector network analyzer (VNA) (**Figure 1B, C**). Building on the S-parameter measurements, the EMI shielding efficiency of several composites was evaluated in the frequency range of 8-12 GHz (X-band). The composite ink can be deposited over various substrates through roll-to-roll printing, laser patterning, and circuit fabrication (**Figure 1D-F**). We also show the application of the composite in wearable monitoring through blocking Bluetooth signals (**Figure 1G**), and as thermal interface materials (TIMs) (**Figure 1H**). The developed composite exhibits outstanding EMI shielding effectiveness (>75 dB in the X-band frequency range of 8-12 GHz), greatly exceeding conventional commercial shielding materials (approximately 20 dB) and surpassing military shielding requirements (>60 dB), even at 200% strain. In addition, it meets the requirements for medical applications, which range from 60 to 120 dB. This breakthrough represents a significant advancement in EMI shielding. The composite waste collected from used composites and circuits was processed to recover LMs using deep eutectic solvent (DES) (**Figure 1I**).

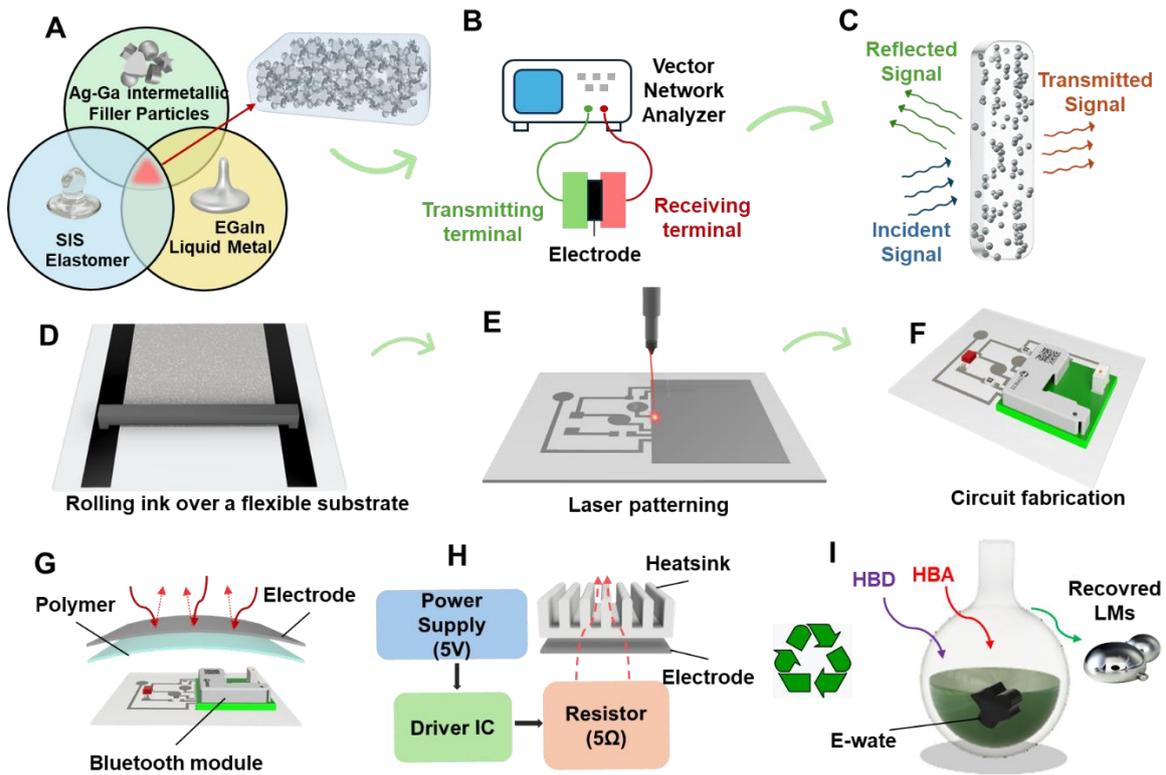

**Figure 1.** A schematic presentation of **(A)** EGaIn- Ag-Ga intermetallic Filler Particles-SIS composite components, **(B)** S-parameters measurement based on a Vector Network Analyzer (VNA), **(C)** VNA receivers measuring the resulting signals from the composite, **(D)** roll-to-roll printing of composite ink over a substrate, **(E)** Laser patterning of the composite, **(F)** Circuit fabrication using the as-fabricated circuit, **(G)** Bluetooth module used to evaluate the performance of the as-fabricated composite as shielding material, **(H)** The as-fabricated composite used as thermal interface material (TIM), **(I)** DES recycling process used for LM recovery, where DESs are made by mixing a hydrogen bond acceptor (HBA) with a hydrogen bond donor (HBD).

**Results and discussion**

To synthesize intermetallic components (IMC) among the Ga-Ag interaction, we exploit galvanic replacement reactions (GRRs). Through GRR, silver nitrate can be reduced on the surface of Ga Liquid metals (LMs). From an electrochemical perspective, the feasibility of galvanic replacement of Ga with $Ag^+$ is attributed to the lower standard reduction potential of Ga (-0.56 V), compared to $Ag^+$ (0.78 V) [30]. We first analyzed exposing a drop of liquid metal to very low molar ratios of 0.1 mM to 5 mM Ag Nitrate. Their corresponding optical images are depicted in **Figure S1**, indicating a formation of Ag nanoparticles/layers over the Ga droplets. As the molar ratio increased, a notable change was observed in the coloration of the LM droplet surfaces, transitioning from nearly black to a goldish hue (**Figure S1**). It indicates that Ga particles comprise a liquid Ga core encased by native surface oxides [31]. This distinctive composition imparts adhesive properties to the Ga particles, stemming from the synergistic influence of the surrounding surface oxides and the pliability of the liquid core [32]. **Videos S1-S2** show the live demonstration of electrochemical reactions that occur in the solution. In fact, the progression of the reaction is likely attributed to the presence of a thin metal oxide layer and the significant disparity in standard reduction potentials between the electron donor (Ga LM) and the electron acceptor (silver ions). Consequently, a thermodynamically spontaneous reaction occurs on the Ga oxide surface, driven by the difference in the reduction potential against the standard hydrogen electrode (SHE), of $Ga^{3+}/Ga$ (-0.529 V) and $Ag^+/Ag$ (0.799 V) [30]. This leads to the formation and subsequent attachment of silver nanocrystals onto the GaLM oxide surface. The reaction can be represented as Ga (s) + $3Ag(aq)^+ \rightarrow Ga\ (aq)^{3+} + 3Ag(s)$, with a standard reduction potential of 1.360 V.

To increase the rate of GRR, the previous samples — containing 0.8 mg of molten gallium in solutions with 0.1 mM, 0.5 mM, 1 mM, and 5 mM concentrations of $AgNO_3$ in Milli-Q (MQ) water — underwent ultrasonic irradiation to induce particle precipitation, schematically shown in **Figure 2A** and **Figure S2**. Subsequently, unreacted nitrate was removed through solution washing, followed by drying of the resulting samples. During the ultrasonic process, the Ga droplets would further reduce in nano or submicron-size particles covered by Ag NPs. **Figure S3** shows the SEM images of the 5 mM samples, where the submicron particles with heterogeneous distribution of the particles are observable, which is confirmed by mapping analysis (**Figure S3).** The X-ray diffraction (XRD) pattern of the 5 mM sample reveals the presence of gallium oxides, including $Ga_2O_3$ and GaOOH, along with small peaks corresponding to Ag (**Figure S3)**. The formation of such phases is confirmed by their corresponding reference patterns shown in **Figure S4**. Such crystals of Ga, i.e. $Ga_2O_3$ and GaOOH, are known to form on the gallium surface through a reaction between gallium, OH radicals, and oxygen in MQ water, as represented by the equation [33]:

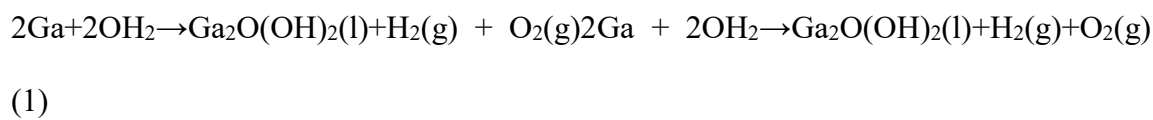

$$2Ga + 2OH_2 \rightarrow Ga_2O(OH)_2(l) + H_2(g) + O_2(g)$$

(1)

Previous studies have demonstrated that the high thermal energy generated during sonication contributes significantly to the rapid formation of GaOOH crystals [20-21]. This phenomenon occurs due to the increased molecular decomposition of water into hydrogen and hydroxyl radicals, acting as oxidizing agents. This principle underscores the mechanism driving the formation of GaOOH crystals on gallium surfaces under sonication conditions. To address Ga oxidation and prevent the formation of such phases, we studied GRR in different solution concentrations. The STEM images of the samples at 0.5 M indicated aggregations of smaller particles on the surface of Ga particles (**Figure 2B.** We should mention that, to preserve the surface particles, the ultrasonicated synthesis particles were directly drop-cast onto carbon tape, allowed to dry, and then analyzed via

STEM. STEM analysis was performed on both the surface and cross-section of the samples synthesized at 0.5 M to investigate the formation of the Ag shell on Ga particles (**Figure 2B, and C**). A clean cross-section was prepared using sample preparation followed by ion milling. High-resolution STEM imaging, utilizing InLens detectors, clearly revealed the core-shell architecture. Additionally, hair-like structures observed on the shell surface provide insight into the Ag growth mechanism around Ga (**Figure 2B-C**). This is the first time this growth mechanism has been visualized, enabled by high magnification and advanced detection. Further STEM analysis confirmed the presence of an Ag shell surrounding the Ga particles. The cross-sectional images indicate that Ga particles are covered by GaAg and Ag shells (**Figure 2C**). **Figure S5-A** shows cross-sectional images of Ag@Ga particles at different magnifications. The images reveal that Ga particles (grayish) are coated with the Ag particles (having a brighter shell) or Ga-Ag biphasic layers (white grayish), where the Ga particles appear grayer. Some Ga liquid metal particles are observed in non-coated states, likely due to ion-milling during sample preparation for SEM analysis. This was further confirmed by taking EDX spectra, which presented different compositions in various zones, suggesting the formation of either GaAg intermetallic compounds or Ag shells on the Ga particle surface**.** The particles were distributed in the range of ~100-1700 nm with an average size of ~548 nm (**Figure 2C**).

Further analysis was performed on the dried powder particles of the as-synthesized samples at 0.5 M. As depicted in **Figure 2D**, the SEM images indicate that the particles became more aggregated due to the malleable behavior of the as-synthesized particles. This behavior results in a broader particle size distribution, ranging from 1 to 12 μm, as shown in **Figure S5-B**. The XRD pattern reveals distinctive features: for samples prepared with Ag nitrate solutions at concentrations of 0.5 M and 0.1 M, Ag is prominent, exhibiting intense diffraction planes such as (111), (200), (220), and (311) at 2θ = 38.38, 46.14, 67.44, and 77.70 [34], along with minor traces of the $Ga_xAg_x$ phase e.g., $GaAg_2$, $GaAg$, and $Ga_{0.5}Ag_{1.5}$ on the surface of the Ga particles (**Figure S4** as well). Increasing the silver nitrate concentration from 5 mM to 0.5 M not only prevents the unwanted Ga oxide phase formation but also allows for considerably faster synthesis of biphasic powder.

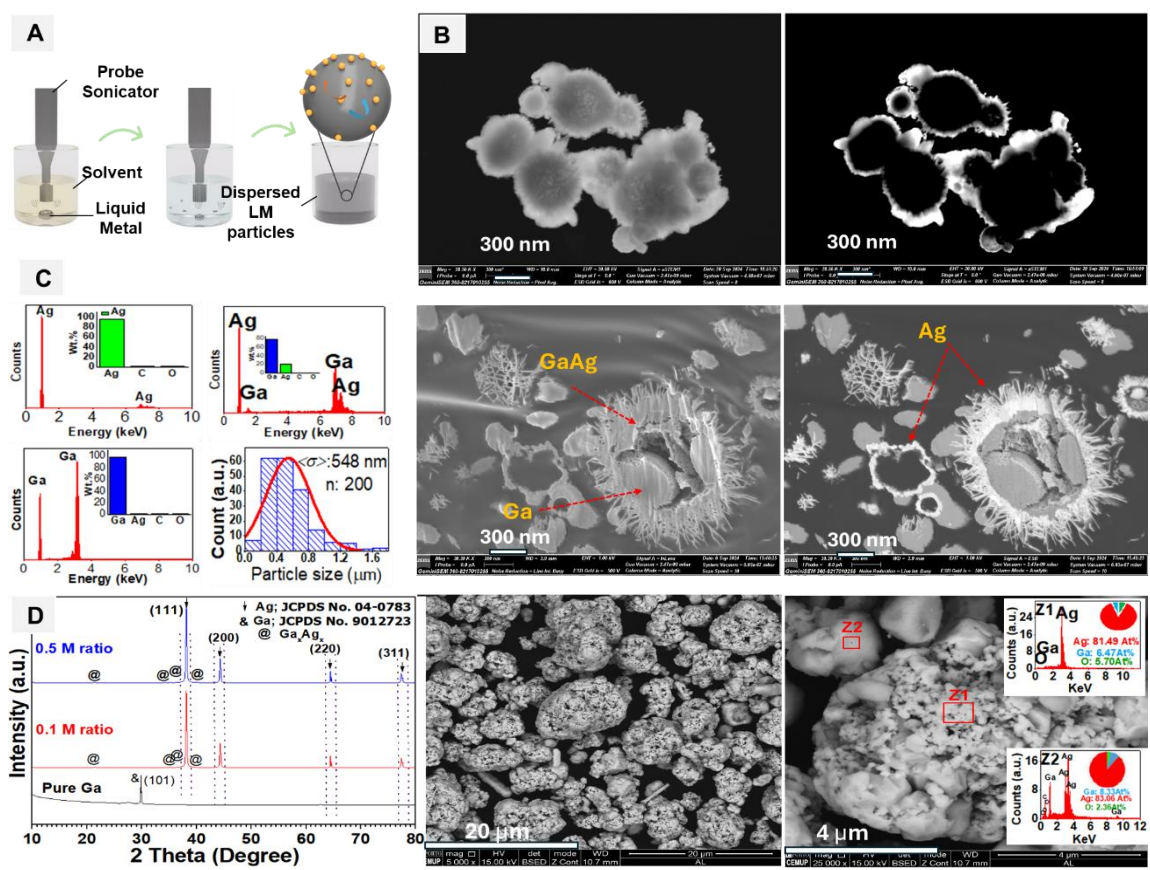

**Figure 2**. **(A)** Schematic representation of the ultrasonic Galvanic Replacement Reaction (GRR) in the presence of silver nitrate aqueous solution. **(B)** STEM images of the 0.5 M silver nitrate samples, including surface **(B)** and cross-sectional **(C)** views, accompanied by EDX spectra and particle size histogram. **(D)** SEM images of the collected powder particles, along with EDX spectra, particle size histogram, and XRD patterns.

**Electrical and Electromechanical behavior of the as-synthesized intermetallic powders:** We first synthesized a composite by incorporating (Ag-Ga) intermetallic filler particles – produced from 0.5 M silver nitrate concentration- into SIS polymer. It is worth mentioning that a mixture of Ga with SIS alone is not sinter-free conductive and requires either mechanical strain or sintering to become electrically conductive. Thus, we incorporated these biphasic filler particles into SIS polymer with a ratio of 1.65 g to 0.2 g. The developed composite could be digitally printed using a simple extrusion nozzle on various substrates including hyperelastic SIS block copolymer and textiles as demonstrated in **Video S3** and **Figure S6**. **Figure 3A** shows the surface SEM images along with their corresponding elemental mapping (**Figure 3A, i-ii). Figure 3A, iii-iv,** shows the cross-section of the same, demonstrating the distribution of the aggregated particles within the SIS polymer matrix. Subsequently, dogbone-shaped samples of the composite (**Figure S7**) underwent electromechanical strain at 30%, 50%, and 100% for 10 cycles. **Figure 3A (v)** showcases the electromechanical behavior of the composites, registering an initial resistance ($R_0$) of approximately 20 $\Omega$ for a sample 50 mm long and 5 mm wide. This is a significant improvement in the electromechanical behavior of the as-fabricated composite compared to the conventional mixture of SIS and Ga liquid metals, which is often reported as nonconductive or requiring mechanical sintering to become conductive [19]. We hypothesize that this is due to the rigid shell around the Ga particles in the SIS polymer matrix, which, similar to solid microfillers and unlike the liquid filler, can make electrical contact with adjacent particles within the polymer matrix. Moreover, the gallium oxide shell that forms around the gallium particles is a semiconductor, while AgGa IMCs are conductive. For comparison, the SEM images of the GAIMCs-SIS and Ga-SIS composites are shown in **Figure S8**. They indicate that the intermetallic compounds of Ga-Ag are connected to each other within the matrix in the former composite. This is one reason we increased the silver nitrate concentration from 5 mM to 0.5 M, which prevents the oxidation of Ga particles and further improves electrical conductivity.

When subjected to 30%-100% mechanical strain, the samples exhibited low electromechanical coupling, as seen in **Figure 3A(v)**. We hypothesize that the liquid metal (LM) within the biphasic particles contributes to the stable behavior under mechanical strain. However, as visible from the SEM images, despite the well-distributed particles, the composite appears to have low contact points and poor percolation. On the other hand, compared to the 1.65g Ag flake-0.2g SIS composite [19], the 1.65g GAIMCs-0.2g SIS composite shows an approximately 60% improvement in electrical conductivity. This suggests not only an enhancement in the electromechanical properties but also opens the door to using low-cost filler particles instead of expensive Ag flake particles in flexible circuits and electronic systems.

We also explored blending Eutectic Gallium Indium (EGaIn) LM into the composite. **Figure 3B, (i-iv)** presents surface and cross-sectional SEM images of the EGaIn-SIS-IMCs composites. The inclusion of 3g EGaIn further improved both initial resistance and electromechanical properties reaching to an initial resistance ($R_0$) of $3 \pm 0.3$ Ω (**Figure 3B, v**). Furthermore, the electromechanical behavior is studied for different compositions, as shown in **Figure S9**. Under a 100% strain, the resistance ($R_u$) decreases to $0.80 \pm 0.04$ Ω, which is due to the rupture of Gallium oxide around the LM droplets, further improving the percolation between the microfillers. In comparison with SIS-Ag flake-LM composites reported in our previous work [19], utilizing Ga LM-coated intermetallic compounds instead of Ag flake filler particles in the composite improved the electromechanical behavior. This is probably due to the substitution of rigid Ag flakes with softer gallium-based core-shell particles. Notably, the LM Filler/Ag ratio was as well reduced from 4:1 to 2:1 compared to the previous work. This suggests that the percolation of biphasic Ag-coated Ga particles is better than Ag flakes previously used. The electromechanical behavior of the samples subjected to 1000 cycles for 100% is shown in **Figure S10**. We therefore selected this composite for further analysis, as it offers the best combination of high electrical conductivity, low electromechanical coupling, and digital printability.

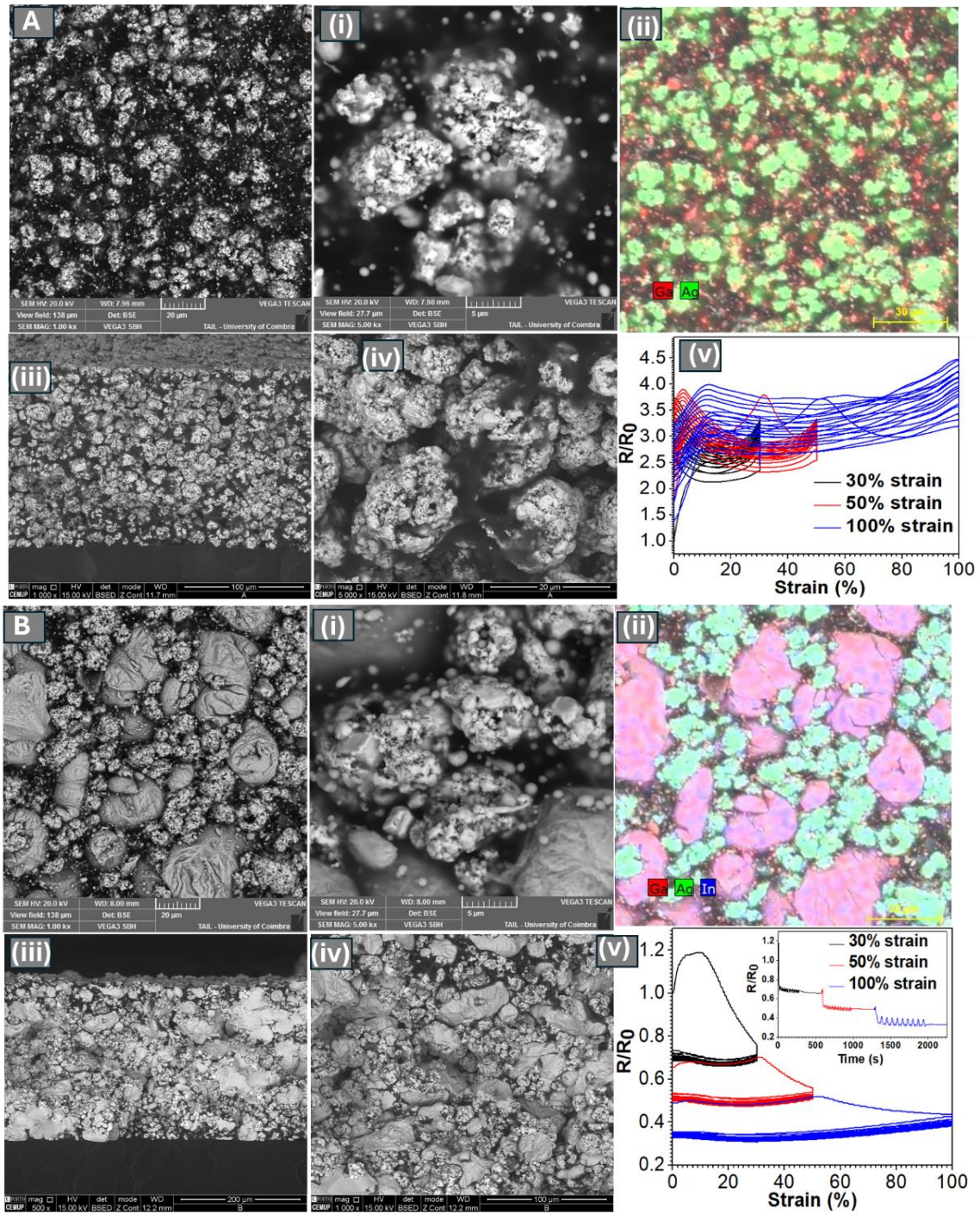

**Figure 3. (A)** SEM images of the surface of the SIS-GAIMCs composite, including (i) high magnification, (ii) mapping, (iii) cross-sectional SEM images, and (iv) high magnification, along with electromechanical behavior (v). **(B)** SEM images of the surface of the SIS-GAIMCs-EGaIn composite, including (i) high magnification, (ii) mapping, (iii) cross-sectional SEM images, and (iv) high magnification, along with electromechanical behavior (v).

The electromagnetic interference (EMI) shielding efficacy (SE) of the samples is assessed using a vector network analyzer (VNA: Rohde & Schwarz ZVB20) across the frequency spectrum from 8 to 12 GHz, using a setup shown in **Figure S11**. This evaluation involves post-processing the measured scattering parameters (S11, S12, S21, and S22). The SIS polymer substrate exhibits zero SE due to its non-conductive nature (**Figure 4A**). Printing Ga LM directly onto the SIS substrate yields an SE value of approximately 10 dB. However, when printing SIS-GAIMCs composite ink over the SIS substrate, the SE exceeds 48 dB in the X-band (**Figure 4A**), meeting commercial standards (>30 dB). We hypothesize that the increase in SE is attributed to the biphasic structure of the as-fabricated composite, which results in a significant increase in the conductive surface area. The microstructural behavior of composites with different morphology and structure of the filler particles facilitates/improves effective multiple reflections and scatterings of electromagnetic (EM) waves within the composite, thereby enhancing its SE performance through effectively attenuating the waves and extending transmission paths [35]. Minor amplitude fluctuations are noticeable, primarily attributed to localized losses occurring during the attenuation of electromagnetic waves [36].

It is indeed crucial to emphasize the significant impact of electrical conductivity on the shielding performance of materials. Therefore, we anticipated that by incorporating EGaIn LM droplets into the SIS-GAIMCs composite, the shielding effectiveness (SE) would increase. This expectation stems from the synergistic effects of the type and structure of the intermetallic filler particles, along with the presence of conductive EGaIn LM particles with fluidic properties within the composite that improve the interface contact between the particles. This contributes to the formation of an exceptional internal conductive network, as illustrated in **Figure 4B,** and is also shown in **Figures S12 and S13** based on the SEM images. Consequently, the SE of the LM-containing composite reaches an exceptional value of 55 dB in the X-band in the unstretched state. The interconnected biphasic structure and morphology of the conductive filler particles and EGaIn LM play a crucial role in attenuating incident electromagnetic─ increasing the spread paths of electromagnetic waves─ through scattering, reflecting, and absorbing mechanisms within the composite. These effects primarily originate from mobile charge carriers and electric dipoles, respectively [37]. The diverse interfaces between the filler particles and LM particles within the SIS matrix atop the SIS substrate have the capability to accumulate free charges, while strategically positioned defects on the surface serve as efficient electric dipoles. These combined factors synergistically contribute to polarization relaxation loss when exposed to alternating electromagnetic fields [38-39]. Furthermore, we hypothesize that intermetallic phases, such as $Ag_2Ga$ and $AgIn_2$, induce further reflections and absorptions due to impedance mismatches at interfaces, confining waves inside the composite and converting them into heat. Scattering at the composite's surface reduces the intensity of reflected waves, contrasting with flat metal surfaces. Absorbed waves dissipate as heat through internal conductive networks via the Joule effect [40-41]. Further analysis, taking into account the reflection coefficient (R ≈ 97%) and absorption coefficient (A ≈ 3%), provides insights into the shielding mechanism. Despite the shielding effectiveness ratio (SER) being lower than the shielding effectiveness of absorption (SEA) for the highly conductive LM coating, the higher value of R compared to A emphasizes that the EMI shielding mechanism is primarily governed by reflection. Similar to our findings, it has been reported that the inclusion of Galinstan liquid metal (LM) in conductive MXene composites led to a notable increase in shielding effectiveness (SE) from 20 dB to 62 dB, which is attributed to the rise in electrical conductivity and the composite density [42]. Similarly, adding EGaIn to cellulose-LM composites resulted in an increase in the SE from 21.5 dB to 33.14 dB [12]. Works referenced in the Ashby diagram (**Figure 4C**) suggest that incorporating liquid metals (LMs) into various composites enhances electromagnetic interference (EMI) SE efficiency [43-46].

It is interesting to note that, upon stretching the sample to 100% strain, the EMI shielding effectiveness (SE) increased to approximately 54.5 dB. Further stretching to 200% strain resulted in an SE exceeding 74.8 dB in the X-band, meeting military shielding requirements (>60 dB), and medical device requirements (>60 dB and < 120 dB [47-48]. The significant enhancement (~19 dB) in EMI shielding performance with increasing strain up to 200% is attributed to the capability to establish more extensive conducting networks through the utilization of conductive fillers. By contrast, particle-filled composites such as carbon nanotube-doped thermoplastic polyurethane experience a dramatic decrease of over 60% in EMI shielding effectiveness (SE) when subjected to mechanical strain (**Figure 4C** and **Table S1**). However, further stretching to 300% strain causes the particles to lose their integrity and reduce the SE performance by about 8% dB due to the increase in resistance with the increase in the applied strain, which is shown in **Figure S14.** In addition to **Table S1**, several other studies highlight the impact of liquid metal (LM) content, filler particles, composite types, and structural design on the shielding efficiency of the materials. The MXene/liquid metal (MLM) plasticine, made using a solvent-assisted dispersion (SAD) method, achieves exceptional shielding with 105 dB [49]. Porifera-inspired MXene foams, featuring MXene/reduced graphene oxide and silver, deliver 53.5 dB shielding at a low density of 38.9 mg/cm³ [50]. Polydimethylsiloxane-packaged films with nitrogen-doped MXene and silver nanowires show 73.2 dB shielding [51]. The LMCPG hydrogel composite reaches 75.69 dB shielding with 8 grams of LM content [52], while the super-stretchable liquid metal foamed elastomer composite (LMF-EC) increases shielding from 57 dB to 85 dB at 400% strain [53]. In another study, a 3D LM network polymer composite shows EMI shielding of 55.7 dB at 0% strain and 63.9 dB at 400% strain [54], and a liquid metal composite with $GaIn_{24.5}$ and Ni achieves over 75 dB shielding across 100 kHz to 18 GHz [55]. These innovations highlight the growing potential of LM composites in advanced electronics and wearable technology.

Additionally, the overall Electromagnetic Interference (EMI) shielding effectiveness (SE) of shielding materials depends not only on the electrical conductivity but also on the thickness of the final composite, according to SE = 9.68 $t$ $(\sigma\omega\mu/2)^{1/2}$, where $t$ is the thickness of the composite, $\sigma$ is the electrical conductivity, $\omega$ is the angular frequency, and μ is the relative magnetic permeability [56]. Therefore, we measured the EMI shielding efficiency of the samples prepared in this study, with a thickness of ~170 μm as well (3D Profilometer images shown in **Figure 4D**). We compared our results with previous studies on composites of varying thicknesses [57-62], as shown in **Figure 4E** and **Table S2**. The enhancement in EMI SE at a reduced thickness of about 170 μm can be attributed to the synergistic effect of the $Ga_xAg_x$-coated Ga particles and EGaIn LM particles in the composite formation, allowing us to achieve higher SE.

To evaluate the real-world viability of the as-fabricated circuits, we conducted the following experiment. A mobile phone was used to send a signal to a circuit that activates an LED via Bluetooth (**Figure 4F** and **Figure S15**). When not covered, or covered with an SIS substrate, the signal could still pass through (**Figure 4G**). However, when the circuit was covered with the as-fabricated composite, there was significant attenuation of Bluetooth signals, with the signal strength dropping from 46 to 70. Although shielding was not complete, the signal could still pass through when the mobile phone was a few centimetres away.

To evaluate the real-world viability of the as-fabricated circuits, we conducted the following experiment. A mobile phone was used to send a signal to a circuit that turns on an LED via Bluetooth (**Figure 4F and Figure S15**). When not covered, or covered with the SIS substrate, the signal could still pass through (**Figure 4G**). On the other hand, when the circuit was covered with the as-fabricated composite, there was a significant blockage of Bluetooth signals, with the signal strength dropping from 46 to 70 counts ("count" refers to the strength of the Bluetooth signals), even though shielding was not complete, and the signal could pass through when the mobile phone was a few cm away. In stark contrast, when the composite was stretched to 200% of its original size and placed above the operating circuit, the phone signal was entirely blocked. Data recorded from Oscilloscope software ─ is used to monitor and analyze wireless signals to identify and prevent unauthorized Bluetooth communications─ (**Figure S16, and Figure 4H**), confirms the previously mentioned data. This underscores that deformation up to 200% strain not only does not reduce the shielding function but even enhances it. Therefore, the fabricated composite can be considered a robust solution for the electromagnetic protection of human organs for wearable monitoring devices, or in general human shielding wearables.

**Thermal interface test setup:** Based on previous studies on the use of liquid metals as thermal interface materials [63-64], we developed a setup that includes a general-purpose DC motor driver (TB6612FNG, Pololu) (**Figure 4I**), which serves as the driving current circuit for a 5.0 Ω, 10W power resistor. The heat generated by the driver chip and resistor is regulated by adjusting the input voltage (0 - 13.5 V) using a power supply (6302D, Topward). An aluminum heatsink was employed to assess heat dissipation for both the driver IC and the resistor during testing (**Video S4** and **Figure S17**). The results indicated that without the thermal interface material (TIM) between the heat sink and resistor, the resistor temperature reached approximately 90 °C after 45 minutes (**Figure 4I**). In contrast, using commercial composite ink as TIM resulted in a recorded temperature of 65.8 °C. Notably, our produced composite ink provided a slight further reduction, bringing the temperature down to 55.9 °C. This demonstrates an improvement in the performance of our composite ink compared to the commercial ink. The thermal performance of the EGaIn-Ag@Ga-SIS composite as a thermal interface material (TIM) was evaluated using an electrical power supply to generate heat. A simple experimental setup (**Figure S18**) was employed, connecting the power supply to a resistive heating element. The heat output was controlled by adjusting the voltage and current, with the temperature set at 50 °C to simulate typical CPU heating conditions.

A 1 mm-thick composite with a 5 cm² surface area was positioned and sandwiched between the heater and a heat sink, and thermocouples were used to measure the temperature difference across the composite. The measured temperature difference was 15 °C. Applying Fourier's law, the thermal conductivity was calculated to be 65.2 W/m·K, highlighting the composite's excellent heat dissipation properties.

**Recycling:** To extract liquid metal (LM) droplets from the composite, we first separated the composite ink from the substrate using acetone (**Video S5**). We then used a deep eutectic solvent (DES) (Choline Chloride (ChCl): Urea with a molar ratio of 1:2) due to its cost-effectiveness, biodegradability, and eco-friendly properties [65]. The DES was prepared by heating the ChCl: Urea mixture to 115 °C until a homogeneous solvent was obtained. Subsequently, the collected waste materials were immersed in the solvent and stirred for 30 minutes (**Figure 4J**). While LM droplets were successfully obtained, some debris from the waste was also observed. However, subjecting this debris to 10 minutes of mechanical sonication did not yield additional LM. In contrast to our previous work, where choline chloride and glycerol were mixed in a 1:2 molar ratio [15] for recovering LM particles from Liquid Metal-Carbon black-SIS polymers, and choline chloride and oxalic acid (OA) were used in a 1:2 ratio [66], achieving a LM recovery rate of approximately 98%, the recovery rate in this study was lower, around 88%. This decrease in recovery efficiency may be associated with the formation of InAg$_2$ intermetallic compounds, which hinder the recovery of some LM particles. We fabricated a circuit using LM particles obtained from recycled e-waste and used LEDs to demonstrate the proper performance of the circuit made from Liquid Metal-As-synthesized filler particles-SIS polymer composites, as shown in **Video S6**.

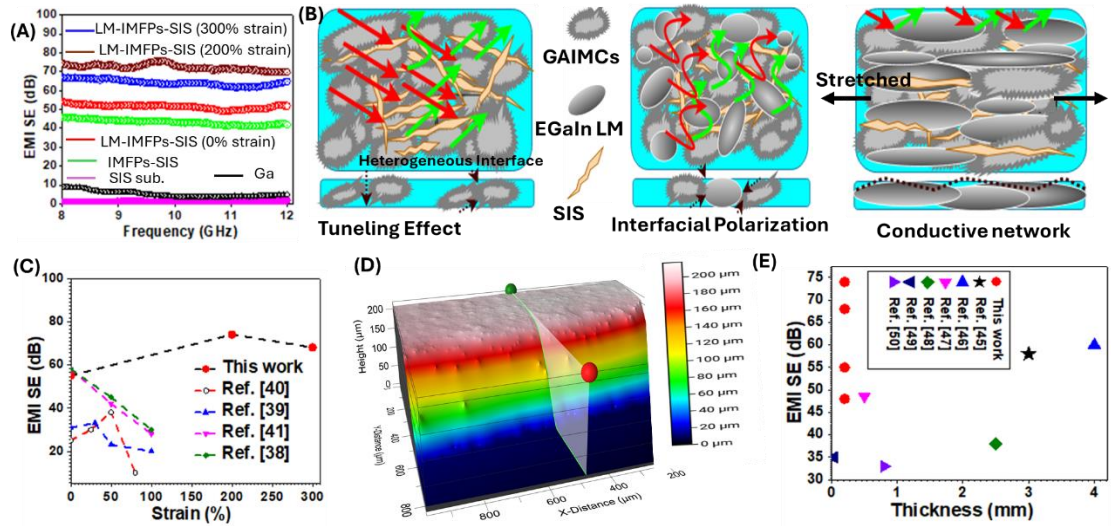

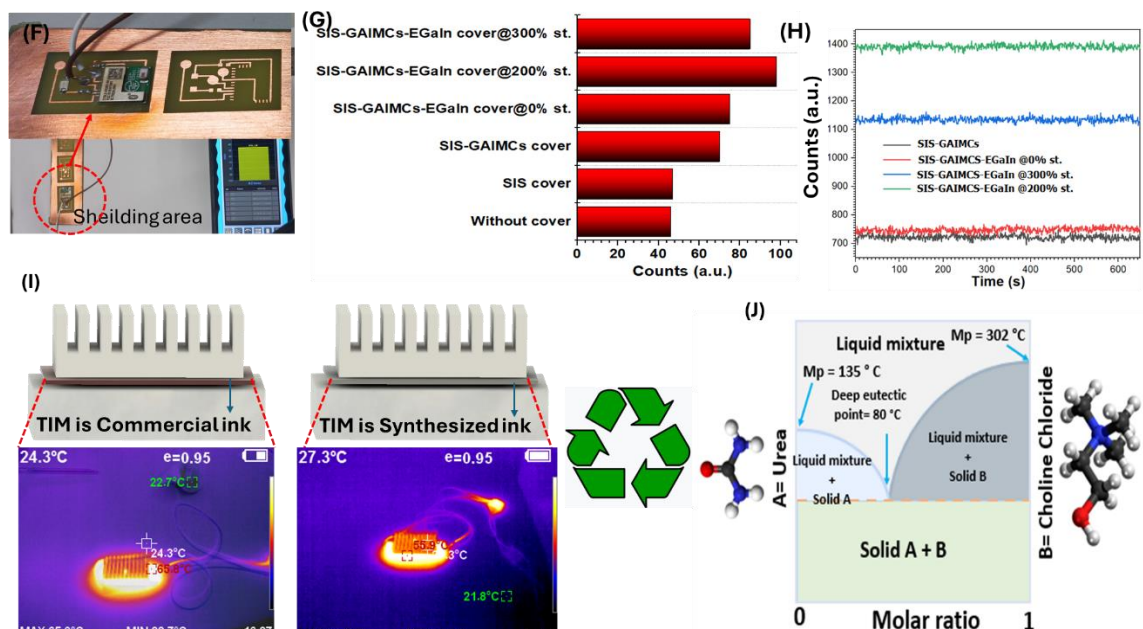

**Figure 4.** (A) EMI shielding efficiency of the composite in different states, (B) Schematic representation of the particle distribution within the composites, (C) Comparison of literature results with this study under different strains, (D) 3D profilometer image of the SIS-GAIMCs composite, (E) EMI SE at different thicknesses, (F) The designed circuit with a PCB used for Bluetooth blocking, and the obtained results (G). (H) Results from the oscilloscope program evaluating the Bluetooth-blocking characteristics of the composites, (I) Captured images of the TIMs by thermal camera, and (J) DES employed for LM extraction.

## Conclusion

We have demonstrated the successful synthesis and utilization of printable, stretchable liquid metal intermetallic composites for high-performance electromagnetic interference (EMI) shielding in soft electronics. By leveraging galvanic replacement reactions (GRRs), we coated gallium (Ga) liquid metal particles with silver (Ag) intermetallic compounds, enhancing their conductivity and compatibility with polymer matrices. Our findings indicate that a conductive composite made by blending the Ag-coated Ga particles has better electrical and electromechanical properties compared to a composite with Ag flakes. Furthermore, blending LM into the developed composite results in further improvements in electrical conductivity, EM shielding, and electromechanical coupling.

Microwave absorption studies revealed exceptional EMI shielding effectiveness (SE) values exceeding 75 dB in the X-band frequency range, even under 200% strain, thereby meeting rigorous military and medical device requirements.

Practical wireless communication experiments further underscored the real-world viability of the composites, effectively blocking Bluetooth signals even when stretched. Additionally, our exploration of recyclability using a biodegradable deep eutectic solvent highlights the eco-friendly nature of these materials, contributing to sustainability efforts in electronics manufacturing.

Overall, our findings underscore the potential of printable, stretchable liquid metal intermetallic composites as versatile solutions to address the growing demand for flexible, high-performance EMI shielding in soft electronics. These materials hold promise for applications in wearable devices, bioelectronics, and flexible displays, facilitating advancements in next-generation electronic technologies. Future research directions may include further optimization of composite properties, scalability of manufacturing processes, and exploration of novel applications in emerging electronic devices.

**Experimental**

**Sample Preparation and Characterization**

The polystyrene-block-polyisoprene-block-polystyrene (SIS) polymer was synthesized in a mixture comprising 63 mL methyl acetate, 20 mL cyclohexane, and 17 mL 4-chlorobenzotrifluoride, referred to as GS, as outlined in previous studies [7],[19]. For ink formulation, a specified quantity of the SIS polymer was combined with the synthesized powder in a rotary mixer at 2000 rpm for 3 minutes. Similarly, when incorporating EGain (a mixture of 75.5% Ga and 24.5% In) with the as-synthesized powders and SIS polymer, the same SIS-powder preparation procedure was followed. The resulting samples were then laser-cut into dog-bone shapes (dimensions: 80 mm height, 20.25 mm width, and 4.86 mm necking part width) for electromechanical measurements using an Instron (Model 5943). Morphological and microstructural analyses of the composites were performed using scanning electron microscopy (SEM) equipped with energy-dispersive X-ray (EDX) spectroscopy and mapping (Bruker Nano GmbH Berlin, Germany Esprit 1.9 and Detector type: XFlash 410). Phase and structural analysis were conducted following methods detailed in Ref. [67].

To determine the EMI shielding effectiveness (SE), the complex scattering parameters (S-parameters) of the material samples were measured and subsequently post-processed. The measurements were performed using a two-port transmission approach, with the material samples placed inside a rectangular hollow metallic waveguide, filling the entire cross section of the waveguide. A VNA (Rohde & Schwarz ZVB20) was utilized to measure the amplitude and phase of the reflection (S11 and S22) and transmission coefficients (S12 and S21). It is noteworthy that for homogeneous material samples, S11=S22 [72], and for reciprocal materials, S21=S12 [68]. To ensure accurate measurement of the S-parameters, the VNA was initially calibrated using a full two-port calibration method. The scattering properties of the material samples were measured in the X-band frequency range (8–12 GHz).

The EMI shielding effectiveness (SE) was then calculated from the S-parameters as follows [69]:

$$\text{SER} = -10 \log_{10}(1 - |\text{S11}|^2) \qquad\qquad (2)$$

$$\text{SEA} = -10 \log_{10}\left(\frac{|\text{S21}|^2}{1 - |\text{S11}|^2}\right) \qquad\qquad (3)$$

$$\text{SE} \approx \text{SER} + \text{SEA} = -10 \log_{10}(|\text{S21}|^2) \qquad\qquad (4)$$

Here, SER and SEA represent the reflection and absorption components, respectively, of the shielding effectiveness (SE). The shielding efficiency component associated with multiple reflections (SEM) was neglected.

**Composite and ink fabrication:**

The procedure used was the same as reported in refs [7],[19].

**Shielding test setup**

A circuit featuring wireless connectivity, comprising a Bluetooth module (CYBLE-022001-00, Cypress), was utilized for conducting the shielding tests. A custom firmware running on the module was responsible for measuring a proximity sensor and transmitting the data to a computer via Bluetooth communication.

**The Supporting Information**

Ga particles in $AgNO_3$ solution at different molar ratios, with optical microscopy images (Figure S1); schematic of ultrasonic irradiation for accelerating GRR (Figure S2); SEM and XRD of 5 mM $AgNO_3$ sample showing submicron particles and gallium oxides (Figure S3); reference XRD patterns (Figure S4); Ag-Ga biphasic particle distribution and aggregation analysis (Figure S5); digital printing of composite ink (Figure S6); dimensions of dogbone specimen and printed ink for electromechanical tests (Figure S7); SEM images showing particle connectivity in Ga-SIS and SIS-GAIMCs composites (Figure S8); electromechanical behavior of SIS-GAIMCs-EGaIn composite under strain (Figures S9-S10); EMI shielding test setup (Figure S11); SEM and electromechanical behavior of samples under strain (Figures S12-S14); Bluetooth-blocking circuit and oscilloscope program (Figures S15-S16); TIM testing setup (Figures S17-S18).

Comparison of EMI SE stretchable composites with this study (Tables S1-S2).

GRR demonstration in AgNO₃ solution with Ga (Video S1) (MP4); effect of AgNO₃ ratio on Ag@Ga formation via GRR (Video S2) (MP4); digital printing of composite on various substrates (Video S3) (MP4); thermal interface test setup (Video S4) (MP4); recycling LM droplets using DES (Video S5) (MP4); fabricated circuit demonstrating LED performance with recycled LM particles (Video S6) (MP4).

**Acknowledgements:**

This research was supported by the Foundation for Science and Technology (FCT) of Portugal under the CMU-Portugal project WoW (Reference No: 45913), and Dermotronics (PTDC/EEIROB/31784/2017), funded by the EU Structural and Investment Funds (FEEI) through an operational program of the Center Region. This work was also supported by the European Research Council, ERC project Liquid3D, grant number 101045072. Additional funding was provided by Add. Additive (POCI-01-0247-FEDER-024533), financed by Regional Development Funds (FEDER), through the Competitiveness and Internationalization Operational Program (POCI). We gratefully acknowledge access to instruments at the TAIL-UC facility, supported by the QREN-Mais Centro program ICT_2009_02_012_1890. T.A.M. acknowledges FCT for research financial support with reference CEECIND/04530/2017/CP1393/CT0004 under the CEEC Individual 2017, and IT-Coimbra for the contract as an assistant researcher with reference CT/No. 004/2019-F00069.

**Conflict of Interest:**

The authors declare no conflict of interest.

**TOC:**

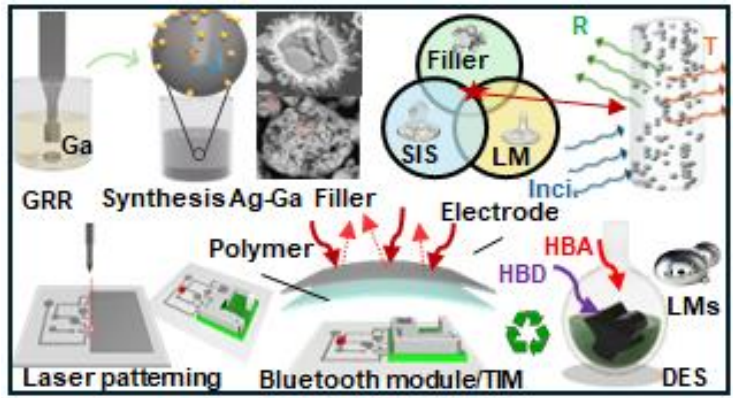